\documentstyle[10pt,epsfig]{article}
\begin{document}
\textwidth 6.75in
\textheight 8.5in

\begin {center}
{\Large Reconciling $\phi$ radiative decays with other data for
$a_0(980)$, $f_0(980)$, $\pi \pi \to KK$
and $\pi \pi \to \eta \eta$}

\vskip 5mm
{D.V.~Bugg\footnote{email address: D.Bugg@rl.ac.uk}},   \\
{Queen Mary, University of London, London E1\,4NS, UK}
\end {center}
\vskip 2.5mm

\begin{abstract}
Data for $\phi \to \gamma (\eta \pi ^0)$ are analysed using the
$KK$ loop model and compared
with parameters of $a_0(980)$ derived from Crystal Barrel data.
The $\eta \pi$ mass spectrum agrees closely and
the absolute normalisation lies just within errors.
However, BES parameters for $f_0(980)$ predict a normalisation for
$\phi \to \gamma (\pi ^0 \pi ^0)$ at least a factor 2 lower than is
observed.
This discrepancy may be eliminated by including constructive
interference between $f_0(980)$ and $\sigma$.
The magnitude required for $\sigma \to KK$ is consistent with data
on $\pi \pi \to KK$.
A dispersion relation analysis by B\" uttiker, Descotes-Genon and
Moussallam of $\pi \pi \to KK$ leads to a similar conclusion.
Data on $\pi \pi \to \eta \eta$ also require decays of $\sigma$ to
$\eta \eta$.
Four sets of $\pi \pi \to KK$ data all require a small but definite
$f_0(1370)$ signal.

\vspace{5mm}

\end{abstract}

\section {Introduction}
Data for $\phi (1020) \to \gamma (\eta \pi ^0)$ and
$\gamma (\pi ^0 \pi ^0)$ from Novosibirsk [1-3] and the KLOE
collaboration at Daphne [4,5] may throw light on $a_0(980)$
and $f_0(980)$, as Achasov and Ivanchenko pointed out in 1989 [6].
A vigorous debate has followed the publication of data.
An extensive list of references is given in a recent
paper of Boglione and Pennington [7].

\begin {figure}  [h]
\begin {center}
\epsfig{file=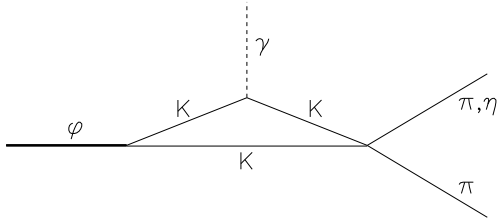,width=10cm}\
\vskip -50mm
\caption{The $KK$ loop model for $\phi$ radiative decays.}
\end {center}
\end {figure}

Fig. 1 shows the model conventionally assumed for radiative
$\phi$ decays.
Both the $\phi \to K^+K^-$ decay and rescattering at the
$f_0$ (or $a_0$) vertex are short-ranged.
In the intermediate state, kaons propagate to larger radii where they
radiate a photon through an electric dipole moment.

These data provide information only
on the lower sides of $a_0(980)$ and $f_0(980)$, i.e. their
coupling to $\eta \pi$ and $\pi \pi$.
There is no direct information on their coupling to $KK$,
except by assuming the $KK$ loop model and using the absolute
normalisation.
For $\gamma (\eta \pi ^0)$, this normalisation depends on the product
$g^2[a_0(980) \to \eta \pi ] \times g^2[a_0(980) \to KK ]$.
In order to test the model, a comparison is made with
parameters of $a_0(980)$ from high statistics Crystal
Barrel data on $\bar pp \to \pi ^0 (\eta \pi ^0)$ [8] and
$\bar pp \to \omega (\eta \pi ^0)$ [9].
There is agreement within the errors, supporting the $KK$ loop model.

The BES collaboration has made an accurate determination of
$f_0(980)$ parameters from $J/\Psi \to \phi \pi ^+ \pi ^-$
and $\phi K^+K^-$ data [10].
Their values of $g^2$ predict a normalisation for
$\phi \to \gamma (\pi ^0 \pi ^0)$ a factor 2
smaller than observed in both Novosibirsk and Daphne data, if
$f_0(980)$ alone is responsible.
This discrepancy is resolved by adding a broad component
coupling to both $\pi \pi$ and $KK$ and interfering constructively
with $f_0(980)$.
It may be parametrised as the high mass tail of the $\sigma$ pole,
although other interpretations are possible, e.g. those
of Au, Morgan and Pennington [11] and of Anisovich, Anisovich and
Sarantsev [12].

This broad component should also appear in $\pi \pi \to KK$.
It will be shown that it can indeed be
accomodated naturally there.
A similar broad component is required to fit data on
$\pi \pi \to \eta \eta$.
A combined fit is made to data on $\pi \pi$ elastic scattering,
$\pi \pi \to KK$ and $\pi \pi \to \eta \eta$. The $\pi \pi$ phase
shifts and elasticities are taken from the re-analysis by
Bugg, Sarantsev and Zou [13] of moments from Cern-Munich data [14].

Formulae are discussed in Section 2.
Readers interested only in results can skip this formalism.
An awkward problem arises in parametrising $\pi \pi \to KK$ near
1 GeV.
The broad component and $f_0(980)$ both go through $90^\circ$ at
$\sim 1$ GeV.
It is difficult to find a parametrisation which accomodates
both resonances while still satisfying unitarity.
The scheme which is adopted here is driven by the features of
the data and is explained in Section 2.

Sections 3 and 4 then report the fits to data on
$\phi \to \gamma (\eta \pi ^0)$ and $\gamma (\pi ^0 \pi ^0)$.
Section 5 concerns the fit to $\pi \pi \to KK$ and $\eta \eta$,
using parameters consistent with $\phi \to \gamma (\pi ^0 \pi ^0)$.
A brief summary is given in Section 6.

\section {Formulae}
\subsection {$\phi$ radiative decays}
Close, Isgur and Kumano [15] give formulae for the mass
distribution of $\pi \pi$ or $\pi \eta$ pairs predicted by the
$KK$ loop model.
It is important to keep track of the absolute normalisation and
also write formulae in a way where the comparison between $\pi ^0
\pi ^0$ and $\pi \eta$ is as simple as possible.
With some trivial re-arrangement of symbols, their eqns. (3.6)
and (3.7) may be written for
$\phi \to \gamma (\pi ^0 \pi ^0)$ as:
\begin {eqnarray}
\frac {d\Gamma }{dm} &=& |I(a,b)|^2 \frac
{\alpha g^2_\phi g^2_{K^+K^-}g^2_{\pi ^0\pi ^0}}{|D(s)|^2}\chi \\
\chi &=& \frac {m_{\pi \pi} k^3_\gamma \rho_{\pi \pi }(s)}
{96\pi ^6m^4_K} \\
k_\gamma &=&\frac {m^2_\phi - s} {2m_\phi }.
\end {eqnarray}
Here $s$ is the invariant mass squared of the $\pi \pi$ pair,
$k_\gamma$ is the photon momentum, $m$ are masses,
$\alpha $ is the fine-structure constant $\simeq 1/137$ and $I(a,b)$ is
a formula for the KK loop integral given by eqns. (3.4) and (3.5) of
Close et al; the quantities $a$ and $b$ are defined there.
The cubic dependence on $k_\gamma$ arises from gauge invariance and is
the familiar $E^3_\gamma$ factor for an E1 transition.
The denominator $D(s)$ in eqn. (1) is written for $f_0(980)$ in the
Flatt\' e form:
\begin {equation}
D(s) = M^2 - s - \frac {i}{16\pi }
[g^2_{\pi \pi }\rho _{\pi \pi }(s)+g^2_{KK}\rho _{KK}(s)];
\end {equation}
the factor $1/16 \pi$ follows the definition used by the Particle Data
Group [16] in their eqn. (38.17), after allowing for a factor $4\pi$
from integrating the S-wave over angles.
Eqn. (4) is the form used by KLOE.
Note that values of $g^2$ in this equation refer to the sum over charge
states:
\begin {eqnarray}
g^2_{\pi \pi } &=& 3g^2_{\pi ^0\pi ^0} =
(3/2)g^2_{\pi ^+\pi ^-} \\ g^2_{KK} &=& 2g^2_{K ^0K ^0} = 2g^2_{K^+K^-}.
\end {eqnarray}
The BES collaboration writes the Flatt\' e form using
\begin {eqnarray}
g'_1 &\equiv &\frac {g^2_{\pi \pi }}{16\pi} \\
g'_2 &\equiv &\frac {g^2_{KK}}{16\pi }.
\end {eqnarray}
Below the $KK$ threshold, $\rho _{KK}(s)$ in eqn. (4) needs to be
continued analytically as $+i\sqrt{4m^2_K/s - 1}$.

Eqn. (3.1) of Close et al. defines $g^2_\phi$ via:
\begin {equation}
\Gamma (\phi \to K^+K^-) = \frac {g^2_\phi }{48\pi m^2_\phi }
(m^2_\phi - 4m^2_{K})^{3/2}.
\end {equation}
The coupling constants $g^2$ for $f_0$ decay are defined to have
dimensions of GeV$^2$, but $g^2_\phi$ is dimensionless and so is
$I(a,b)$. Dimensions of eqns. (1) and
(2) balance between left and right-hand sides; this was not apparent in
the papers of either Close et al or Achasov and Ivanchenko, since
they replace $g^2_{\pi \pi }g^2_{KK}$ by a single $g^2$.
The two powers of $g^2$ appear explicitly in the KLOE publication [5].
The relative normalisation of data on $\phi \to \gamma (\pi ^0 \pi ^0)$
and $\pi \to \gamma (\eta \pi ^0)$ will play an important role:
\begin {equation}
\frac {(d\Gamma /dm)_{\pi ^0 \pi ^0}}{(d\Gamma /dm)_{\eta \pi ^0}} =
\frac
{g^2_{f_0KK}(g^2_{f_0\pi \pi }/3)|D(s)|^2_{a_o}\rho_{\pi \pi }(s)}
{g^2_{a_0KK}g^2_{a_0\pi \eta }|D(s)|^2_{f_o}\rho_{\eta \pi }(s)}.
\end {equation}

\subsection {Identical particles}
At this point, it is necessary to pause and ask whether decays
to $\pi ^0 \pi ^0$ are affected by the fact that they are
identical particles, whereas $\pi ^0 \eta$ are not.
This is a tricky point.
An $I=0$ $f_0$ couples to the isospin combination
$(\pi ^+\pi ^- + \pi ^- \pi ^- -\pi ^0 \pi ^0)/\sqrt {3}$,
where the coefficients are isospin Clebsch-Gordan coefficients.
Experimentally, the $\pi ^+\pi ^-$ integrated  cross section
may be obtained by counting $\pi ^+$ over $4\pi$ solid angle.
There are two amplitudes, one for producing a $\pi ^+$ at angle
$\theta$, say, and the the other for producing a $\pi ^-$
at that angle and an accompanying $\pi ^+$ at angle $(\theta + \pi )$.
Both are in S-waves, so at angle $\theta$ there is a total
amplitude for $\pi ^+$ of $2/\sqrt {3}$.
Integrating over $4\pi$ solid angle, the intensity of observed
$\pi ^+$ is $(4/3)4\pi$. For the $\pi ^0 \pi ^0$ case, there
are likewise two $\pi ^0$ at angles $\theta $ and $(\theta + \pi)$
so the total $\pi ^0$ amplitude is $2\pi ^0/\sqrt {3}$.
In this case, the integration should be only over $2\pi$ solid
angle, so as to avoid counting $\pi ^0$ pairs twice; the integrated
intensity is $(4/3)2\pi$. The total $\pi \pi$ intensity is $8\pi$,
compared with the total $\eta \pi$ intensity of $4\pi$, where there are
no effects for identical particles.

So the $\pi \pi$ amplitude is increased by a factor 2 by the identity
of the two pions. However, it is essential to remember that this
factor 2 appears also in the Breit-Wigner denominator.
The Breit-Wigner amplitude becomes
\begin {equation}
f = \frac {g_{K^+K^-}g_{\pi ^0 \pi ^0}(2/\sqrt {2})(1/\sqrt {3})}
   {M^2 - s - i[2g^2_{\pi \pi}\rho _{\pi \pi } + g^2_{KK}\rho _{KK}]}.
\end {equation}
In the numerator, the factor $2$ allows for the doubling of the
$\pi \pi$ amplitude by identity of the two pions and the factor
$1/\sqrt {2}$ allows for the fact the $\pi ^0$ are to be considered
only over $2\pi$ solid angle.
However, experimentalists do not write the Breit-Wigner amplitude
in this way.
They ignore identical particle effects and write it
in all cases as
\begin {equation}
f = \frac {G_{K^+K^-}G_{\pi \pi }/\sqrt {3}}
   {M^2 - s - i[G^2_{\pi \pi}\rho _{\pi \pi } + G^2_{KK}\rho _{KK}]},
\end {equation}
where $G = g\sqrt {2}$.
It is essential to remember this convention when using Breit-Wigner
parameters deduced in the BES experiment.
The upshot is that in comparing with Kloe data, one should use
eqn. (12) and there is no effect from the identity of the pions in
the final state.

\subsection {How to treat the $\sigma$ pole}
In the KLOE analysis of $\phi \to \gamma (\pi ^0 \pi ^0)$, a
further contribution was included from the $\sigma$ pole.
This requires some discussion.
For elastic $\pi \pi$ scattering, the amplitude may be written
\begin {equation}
f_{el} = \frac {N(s)}{D(s)}
\end {equation}
and there is a zero in $N(s)$ at the Adler point $s \sim 0.5 m^2_\pi$,
just below threshold [17].
The Adler zero is a basic feature of Chiral Perturbation Theory
and figures prominently in the series of papers on
$\sigma$, $\kappa$, $f_0(980)$ and $a_0(980)$ by Oset and
collaborators [18-23].
Data on $J/\Psi \to \omega \sigma$ may be fitted in both
magnitude and phase using
\begin {equation}
f_{prod} \propto \frac {1}{D(s)}
\end {equation}
with the same denominator but a constant numerator [24].
The justification is that left-hand singularities due to
coupling of $\sigma$ to $\omega J/\Psi $ are very distant.
The question is whether the Adler zero is needed in the
amplitude for fitting KLOE data or not.

An interesting intermediate case is the production of the
$\pi\pi$ S-wave in $\Upsilon ' \to \Upsilon \pi \pi $.
Recent BELLE data determine $\pi \pi$ mass spectra for the
transitions $4S \to 1S$, $3S \to 1S$ and $2S \to 1S$ [25].
The first and last follow elastic scattering closely.
The transition $3S \to 1S$ deviates from this and shows a peak
towards the $\pi \pi$ threshold, reminiscent of that observed
in $J/\Psi \to \omega \sigma$, where the Adler zero is absent.
The combined mass of $\Upsilon (3S)$ and $2\pi$ lies close to
the $B\bar B$ threshold, and Moxhay suggests that the decay
proceeds through coupling to $B\bar B$ intermediate states [26].
These data suggest that for small momentum
transfers the Adler zero {\it is} needed in the amplitude
unless there is a flavour change.
From Fig. 1 for $\phi$ radiative decays, it is
plausible that the right-hand vertex should be the same as that in
the on-shell $KK \to \pi \pi$ process, since the kinematics are
similar. It will be shown that this successfully fits the data.

\subsection {$\rho \to \eta \gamma$ and $\pi ^0\gamma$}
There are also amplitudes due to $\phi \to \pi ^0 \rho$,
$\rho \to \eta \gamma$ and $\rho \to \pi ^0 \gamma$.
Formulae for their intensities and interferences with $a_0$ or
$f_0$ are given by Bramon et al. [27] and Achasov and Gubin [28];
a small correction to [28] for the intensity of $\phi \to \rho
\pi$, $\rho \to \eta \gamma$ is made by Achasov and Kiselev [29].
Also eqn. (20) of Ref. [28]
requires an additional factor $(m^2_\phi - s)$ in the numerator;
this correction is made in a second paper of Achasov and Kiselev
[30] in their eqn. (21).

The $\rho \pi$ contribution to $\gamma (\pi ^0 \pi ^0)$ peaks
at $m_{\pi \pi} = 550$ MeV and is small.
Interferences with $f_0(980)$ have rather
little effect on the fit to $\gamma \pi ^0 \pi ^0$.
However, the contribution of $\rho \to \eta \pi ^0$ is quite
significant in $\gamma (\eta \pi ^0)$.

Branching fractions for these processes will now be discussed,
starting with $\rho \to \pi ^0 \gamma$.
Achasov et al. report $\Gamma (\rho \to \gamma \pi ^0) = 73.5 \pm 11$
keV [31].
This agrees with the PDG average for $\Gamma (\rho ^\pm
\to \gamma \pi ^\pm)$ of $68 \pm 7$ keV.
The weighted average is $70 \pm 6$ keV.
The PDG gives an average for $\Gamma (\phi \to \rho \pi + \pi ^+\pi ^-
\pi ^0)/\Gamma _{total} = 0.151 \pm 0.009$.
Dividing this by 3 for neutral final states and assuming it is all
$\rho \pi$, the branching fraction of $\phi \to \rho \pi ^0 \to
\pi ^0 \pi ^0 \gamma$ is $2.4 \times 10^{-5}$.

The experimental groups apply cuts to remove $\phi \to \omega \pi ^0$,
$\omega \to \pi ^0 \gamma$.
It is hard to estimate the fraction of $\rho \pi ^0$ events which this
removes; a sharp cut in the $\rho$ line-shape at
the quoted mass cut leaves $\sim 40\%$ of
$\rho \pi ^0$ events, i.e. a branching fraction of $\phi \to \pi ^0
\pi ^0 \sim 0.96 \times 10 ^{-5}$.
The uncertainty is not serious, since most events lie at low
$\pi ^0 \pi ^0$ masses and have little effect on the
$f_0(980)$ peak.

The arithmetic for the $\rho \pi ^0$ contribution to $\gamma (\eta \pi
^0)$ goes similarly.
The weighted mean of SND [31] and CMD2 [32] values for
$\Gamma (\rho \to \gamma \eta )/\Gamma _{total}$ is
$(2.95 \pm 0.26) \times 10^{-4}$.
Taking the $\phi \to \rho \pi ^0$ branching fraction to be 0.151/3, as
above, and using the KLOE branching fraction $\phi \to \gamma (\eta \pi
^0) = 0.83 \times 10 ^{-4}$, the ratio $(\phi \to \rho
\pi, \rho \to \gamma \eta ) /(\phi \to {\rm all }~\gamma \eta \pi ^0) =
0.18 \pm 0.02$. These events do contribute significantly to the fit;
interference with $\gamma a_0(980)$ is included.

\subsection {Formulae for $\pi \pi \to KK$}
There is a major question how to combine the amplitudes for
$f_0(980)$ and the broad $\sigma$.
Data on $\pi \pi$ elastic scattering from the Cern-Munich experiment
and elsewhere show that phase shifts of
$f_0(980)$ and $\sigma$ add to a good approximation.
Both go through $90^\circ$ near 1 GeV and the phase shift for the
full amplitude passes rapidly through $180^\circ$ just below
1 GeV.
A natural and successful way of describing this situation
is the Dalitz-Tuan prescription [33], where
S-matrices of the two components are multiplied:
\begin {equation}
S_{total} = S_AS_B = \eta _A \eta _B e ^{i(\delta _A + \delta _B)}.
\end {equation}
However, it is not obvious that it is right to multiply elasticities.

An alternative approach is to use the K-matrix.
It is conventional to add K-matrix elements for the two
components [12].
The penalty in this approach is that each K-matrix pole appears
where the total phase goes through $90^\circ$; the $f_0(980)$
is then described as a delicate interference between two poles
displaced substantially from 1 GeV.
In the present work that is inconvenient, since $f_0(980)$ is
the focus of attention in $\phi$ decays; a formula is needed to
parametrise it directly.
Furthermore, it is not obvious that Nature chooses to add K-matrix
elements.

Since $K \propto \tan \delta$, an alternative choice is
\begin {equation}
 K_{total} = \frac {K_A + K_B}{1 - K_A K_B}
\end {equation}
so that phases again add below the $KK$ threshold.
However, this alternative fails to fit $\pi \pi \to KK$ data immediately
above the $KK$ threshold.
This is because inelasticity of the $\sigma$ amplitude grows fairly
slowly above threshold, so eqn. (16) demands a rather elastic amplitude
there.

One cannot add inelasticities of $f_0(980)$ and $\sigma$, since this
leads to values outside the unitary circle.
The approach adopted here is pragmatic.
The Dalitz-Tuan
prescription is simple, fits the phase of elastic scattering
adequately, predicts a reasonable result for the elasticity parameter
$\eta$ and successfully fits the data.
The $\eta$ parameter for the BES $f_0(980)$ drops precipitously to
$\eta = 0.27$ at 1.01 GeV and then rises again (see Fig. 11(a) below).
Since $f_0(980)$ and $\sigma$ are approximately in phase, it seems
likely that the effect of the $\sigma$ component will be to increase
the inelasticity to a value somewhere between
0 and the value for $f_0(980)$ alone.
In practice, multiplying
the S-matrix elements for $S_{11}$ and $S_{22}$ describes the data well
up to 1.2 GeV, obtaining $S_{12}$ from the exact relation
\begin {equation}
2|S_{12}|^2 = 1 + |S_{33}|^2 - |S_{11}|^2 - |S_{22}|^2,
\end {equation} where indices 1,2,3 refer to $\pi \pi$, $KK$ and $\eta
\eta$.
Note, however, that $S_{12}$ itself does not factorise into
$f_0(980)$ and $\sigma$ components.
This is a symptom of extensive
multiple scattering.
It may well be that there is no simple formula
which fully describes strong coupling of the two amplitudes.

Above 1.2 GeV, the appearance of $f_0(1370)$, $f_0(1500)$ and
$f_0(1790)$ makes the treatment via the Dalitz-Tuan prescription
intractable.
There, amplitudes are small enough that unitarity corrections
are less than normalisation errors, and it
is adequate to add amplitudes.
That is the procedure adopted here,
making a smooth join at 1.17 GeV using Lagrange multipliers to ensure
that the magnitude and phase join continuously. For the $\eta \eta $
channel, amplitudes are so small that unitarity plays no significant
role, so one can add amplitudes.

At the higher masses, $f_0(1370)$, $f_0(1500)$
and $f_0(1790)$ overlap significantly.
The overlap makes the phase of the background below each resonance
a sensitive parameter.
In order to achieve a satisfactory fit, it is necesssary to
multiply each amplitude by a fitted phase factor $\exp (i\theta )$ and
add amplitudes.

In summary, multiplying $\eta$ parameters near threshold puts a lower
bound on the intensity 92\% of the unitarity limit, and reduces the
normalisation uncertainty from 50 to 8\%.
At higher masses, the amplitude is fitted freely in terms of known
resonance, ignoring possible multiple scattering from one resonance
to another where they overlap.
This allows maximum freedom to fit $\sigma$ and $f_0(980)$, the main
object of the present work.

An important point is that the $f_0(980)$ signal appears as a sharp
dip in elastic scattering, because phases add.
In $\phi$ radiative decays, it appears as a peak.
In this case, it is appropriate to add amplitudes, since
they are weak and the unitarity limit imposes no significant
constraint; in elastic scattering, the amplitude must follow the
unitarity limit up to the $KK$ threshold.

\subsection {The $4\pi$ channel}
Both $\pi \pi \to KK$ and $\pi \pi \to \eta \eta$ cross sections
drop fairly rapidly with increasing mass, and almost disappear
above 1.9 GeV.
To some extent, this is due to competition from the $4\pi$ channel.
Unfortunately, there are almost no data on the $4\pi$ channel,
so the $s$-dependence of this inelasticity is poorly known.
As in the work of Bugg, Sarantsev and Zou in 1996 [13],
$4\pi$ phase space is parametrised empirically by a formula
close to a Fermi function:
\begin {equation}
\rho _4(s) = \frac
{\sqrt {1 - 16 m^2_\pi/s}}
{1 - \exp [\Lambda (s_0 - s)]}.
\end {equation}
Here $\Lambda = 2.85$ GeV$^{-2}$ and $s_0 = 2.5$ GeV$^2$.
This function approximates closely the onset of $\rho \rho$ and
$\sigma \sigma$ final states.

\subsection {The $\pi \pi $ S-wave}
The propagator of the broad $\sigma$ amplitude is written in eqn. (20)
as the sum of contributions from $\pi \pi$, $KK$, $\eta \eta$ and
$4\pi$, labelled 1--4.
In the first three, the Adler zero is accomodated in eqn. (21) by a
factor
\begin {equation} A(s) = (s - s_A)/(M^2 - s_A),
\end {equation}
where $s_A = (0.41 \pm 0.06) m^2_\pi$.
The factor 0.41 comes from the work of Leutwyler and Colangelo, using
Chiral Perturbation Theory [34].
A form factor $FF_i^2(s) = \exp(-\alpha k^2_i)$ is included into
channels 2 and 3 ($KK$ and $\eta \eta )$ in eqns. (22) and (23).
Note that $FF_i(s)^2$ is multiplying the {\it intensity} of the
$KK$ channel, which is the quantity appearing in the width of the
$\sigma$.
It is required to parametrise the rapid drop of cross sections with
increasing mass.
The same $\alpha$ is used for both channels; $k_i$ is the centre of
mass momentum in each channel.
The propagator is then written:
\begin {eqnarray}
D(s) &=& M^2 - s - g_1^2z_{sub} - m(s) -i[g_1^2\rho _1 +
g_2^2\rho _2 +g_3^2\rho _3 + g_4^2(s)] \\
g_1^2 &=& B_1\exp [-(s - M^2)/B_2]A(s) \\
g_2^2 &=& r_2g_1^2 FF^2_2(s) \\
g_3^2 &=& r_3g_1^2 FF^2_3(s) \\
FF_i(s) &=&\exp (-\alpha _ik_i^2) \\
g_4^2 &=& B_4 \rho _{4\pi }(s)/\rho _{4\pi}(M^2) \\
z(s) &=& \frac {1}{\pi}\left[2 + \rho _1  ln_e
\left( \frac {1 - \rho _1}{1 + \rho _1}\right) \right] \\
z_{sub} &=& z(s) - z(M^2)  \\
m(s) &=& \frac {s - M^2}{\pi}\int \frac {M\Gamma _{tot}(s')ds'}
    {(s' - s)(s' - M^2)}.
\end {eqnarray}
For other resonances, the exponential of eqn. (21)  and the form
factors $FF$ are superfluous because the resonances are narrow;
they are simply omitted, but the dispersive term $m(s)$ is retained.
The  amplitude for $\pi \pi$ elastic scattering is
\begin {eqnarray}
T_{11} &=& 1 + 2iS_{11} \\
S_{11} &=& \eta \exp (2i\delta) = g^2_i \rho_1/D(s).
\end {eqnarray}
In fitting elastic data, S-matrices from $\sigma$, $f_0(980)$,
$f_0(1370)$, $f_0(1500)$ and $f_0(1790)$ are multiplied together.

The function $g_1(s)$ was introduced by Zou and Bugg [35].
It takes a very simple empirical form which
succeeds in fitting the broad S-wave over the whole mass range
from threshold to 1.9 GeV using just  the Adler
zero (taken from other work) and three
fitted parameters $B_1$, $B_2$ and M.
At the $KK$ and $\eta \eta$ thresholds, $r_2 = g^2_{KK}/g^2_{\pi \pi}$
and $r_3=g^2_{\eta \eta}/g^2_{\pi \pi}$.

The motivation for the formalism is to make amplitudes fully
analytic.
This has been achieved below the region of strong $4\pi$ elasticity
by Achasov and Kiselev [30].
Here the opening of the strong $4\pi$ channel is treated analytically
for the first time.
This will be important in demonstrating the presence of $f_0(1370)$;
otherwise, the issue of analyticity
turns out not to be crucial for
present work.
The function $z(s)$ accounts analytically for
the opening of $\pi \pi $ phase space $\rho _1(s)$; it is taken
from the work of Leutwyler and Colangelo [34].
It has the good feature of eliminating the divergence at $s=0$ due to
$\sqrt{1/s}$ in $\rho _1(s) = \sqrt {1 - 4m^2_\pi /s}$.
In eqn. (27), a subtraction is made at $s = M^2$, where the real
part of the S-wave amplitude goes to zero.

\begin {figure} [h]
\begin {center}
\centerline{\hspace{0.2cm}\epsfig{file=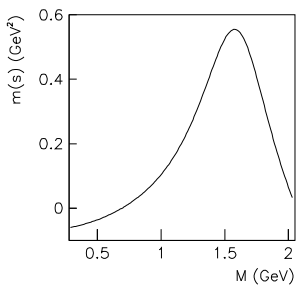,width=5cm}}
\vskip -5mm
\caption{The dispersive term $m(s)$ for the $4\pi$ channel.}
\vskip -5mm

\end {center}
\end {figure}

The term $m(s)$ is a dispersive contribution  to the real part of
the amplitude, allowing for the inelasticity in all channels.
The inelasticity in $4\pi$ is large and gives rise to a
slowly varying dispersive term illustrated in Fig. 2.
It is far too wide to be confused with $f_0(1370)$, $f_0(1500)$
or $f_0(1790)$.
The dispersive term $m(s)$ is also included into the
Breit-Wigner forms for $f_0(1370)$, $f_0(1500)$ and $f_0(1790)$,
again with a subtraction on resonance.
This is important for $f_0(1370)$, where $\Gamma _{4\pi}$ is dominant.

One new result arises concerning the position of the $\sigma$
pole.
Fitting it with the new fully analytic formulae described
above moves its pole position to $(506 \pm 30) - i(238 \pm 30)$
MeV from the BES value of $(541 \pm 39) - i(252 \pm 42)$ MeV [36].
The shift arises from the elimination of the singularity
at $s= 0$ and dispersive corrections to the $\pi \pi$
channel. It is within the experimental errors, which now
go down because of better control over the extrapolation to
the pole.

\begin {figure} [h]
\begin {center}
\epsfig{file=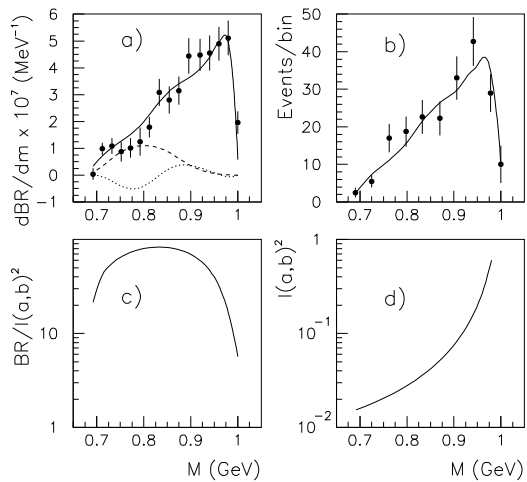,width=8cm}\
\vskip -5mm
\caption{The fit to KLOE data [5] where (a) $\eta \to \gamma \gamma$,
(b) $\eta \to 3\pi$; in (a), dashed and dotted curves show the
$\rho \pi$ contribution and interference with $a_0(980)$;
(c) the energy dependent factor appearing in $d\Gamma /dm$, but
omitting $|I(a,b)|^2$, (d) $|I(a,b)|^2$ itself.}
\end {center}
\end {figure}

\section {Fit to $\phi \to \gamma (\eta \pi ^0)$}
If parameters of $a_0(980)$ are fitted directly to $\phi \to \gamma
\eta \pi ^0$ data of Ref. [5], the problem is that the upper side of the
resonance is not visible, resulting in values of $M$, $g^2_{\eta \pi }$
and $g^2_{KK}$ which are strongly correlated.

Most determinations quoted by the PDG have been obtained by fitting
simple Breit-Wigner line-shapes with constant width to
$\eta \pi$ or $KK$ alone.
These are of no use for present purposes, since they ignore the
cusp at the $KK$ threshold.
A rather precise determination, not quoted by the PDG, was made
in Ref. [8], combining data on $\bar pp \to \omega (\eta \pi ^0)$
and $\bar pp \to (\eta \pi ^0) \pi ^0$ at rest.
Both have high statistics ($280$K events for
$\eta \pi ^0 \pi ^0$) and very low experimental backgrounds.
The first reaction is dominated by $\omega a_0(980)$.
Although a simple Breit-Wigner of constant width was fitted,
the data determine precisely the location of the $a_0$
peak at half-height; the error quoted on the central mass is
$1.23(stat) \pm 0.34(syst)$ MeV and for the width is $0.34(stat) \pm
0.12 (syst)$ MeV. Those constraints go a long way towards breaking the
correlation between couplings to $\eta \pi$ (which dominates the lower
side of the resonance) and $KK$ (which dominates the upper side). In
$\eta \pi ^0 \pi ^0$, the $a_0(980)$ appears very conspicuously in the
Dalitz plot as an `edge' at the $KK$ threshold; at this edge, the phase
turns abruptly by $90^\circ$ in the Argand diagram. The $a_0$
interferes with the broad $\sigma$ in the $\pi \pi$ channel. This
interference determines rather precisely the phase variation of the
$a_0$ as a function of $\eta \pi$ mass. On the other hand, there is
some uncertainty in the parametrisation of the $\sigma$ as a function
of $\pi ^0 \pi ^0$ mass. Taking into account all uncertainties, quoted
parameters are $M = 999 \pm 5$ MeV, $g^2_{\eta \pi } = 221 \pm 20$ MeV
and $r = g^2_{KK}/g^2_{\eta \pi } = 1.16 \pm 0.18$.

The fit to KLOE data with these parameters immediately reproduces
the line-shape accurately.
The fitted normalisation is low by $ 32\%$, which is close to the
combined errors, as follows.
From $g^2_{\eta \pi}$, there is an error of 19\% in the
magnitude predicted for KLOE data, and from $r$ there is a further
error of 15\%. The error quoted by the PDG for
the normalisation of $\phi \to \gamma \eta \pi$ is 6\%.
To achieve agreement in normalisation, it is necessary to
stretch all parameters by $1\sigma$ as follows. For
$a_0$, $g^2_{\eta \pi}$ has been increased to 241 MeV, and
$r = g^2_{KK}/g^2_{\eta \pi } $ to 1.34;
this requires reoptimising $M = 991.5$ MeV. The branching fraction of
$\rho \to \eta \pi$ has been increased to 0.20 and the normalisation of
KLOE data has been decreased by $6\%$. The final fit is shown in Figs
3(a) and (b). Dashed and dotted curves show the intensity of the $\rho
\to \eta \pi$ signal and its interference with $a_0(980)$. A minor
detail is that a form factor $\exp {-\beta k^2_\gamma }$ is included to
allow for the charge radius of the kaon, 0.560 fm. [16].

Figs. 3(c) and (d) illustrate the strong energy-dependent factors
contributing to the intensity. Fig. 3(d) shows $|I(1,b)|^2$
from the $KK$ loop. It falls dramatically as the kaons go off
the mass-shell. Fig. 3(c) illustrates the remaining
factors from $mk^3_\gamma \rho _{\eta \pi }$, eqn. (2).
These factors inflate the lower side of $a_0(980)$.
The agreement with the $a_0$ parameters of Ref. [8]
for both line-shape and normalisation confirms Achasov's $KK$ loop
model at the level of $\sim 15\%$ in amplitude.
That conclusion will be important in considering the $\gamma \pi ^0 \pi
^0$ data.

\begin {figure} [htb]
\begin {center}
\epsfig{file=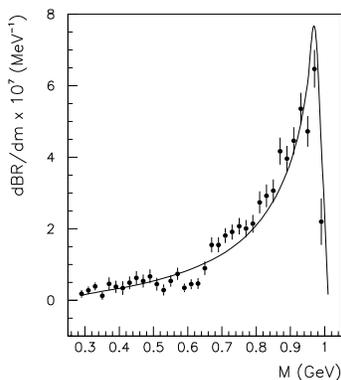,width=6cm}\
\vskip -7mm
\caption{Fit to KLOE $\gamma (\pi ^0 \pi ^0)$ data with $f_0(980)
+ \rho \pi$ when the normalisation is floated.}
\end {center}
\end {figure}

\section {Fit to $\phi \to \gamma (\pi ^0 \pi ^0)$}
KLOE data on $\phi \to \gamma (\pi ^0 \pi ^0)$ [4] can be fitted
fairly well with BES parameters for $f_0(980)$ if the
absolute normalisation is floated, see Fig. 4.
The best fit is obtained by fitting the form factor $FF$ freely and
stretching BES parameters to the limits of their quoted
statistical and systematic errors as follows:
$M = 0.957$ GeV, $g^2_1 = 0.190$ GeV, $g^2_2/g^2_1 = 3.75$.
However, $\chi ^2 = 122$, compared with 60 for the fit shown
later in Fig. 6 including $\sigma \to KK$.
The data are taken from Fig. 4 of the KLOE publication [4] after
subtracting their estimated background; they are not taken
from their Table 5, where the background has been unfolded in
such a way as to accomodate a dip near 520 MeV.

There is however a serious problem with the absolute normalisation.
The fit with $f_0(980)$ alone
has a normalisation at least a factor 2 lower than the data.
The program is identical to that used for $\gamma \eta \pi
^0 $ except for trivial changes in the parameters of the Breit-Wigner
amplitude and the change from the mass of the $\eta$ to the $\pi ^0$.
A direct comparison of the relative normalisation of $\gamma \pi ^0
\pi ^0$ and $\gamma \eta \pi ^0$ has been made using eqn. (10).

A variety of explanations for the normalisation are possible,
but look odd in view of the agreement in normalisation for
$\gamma \eta \pi ^0$ data.
Firstly, Oller remarks that there may be a contact term involving two
$K^0$ in the loop diagram, and a photon radiated from the $\phi$
decay vertex [37].
Secondly, Oset and collaborators have suggested that $\phi$ may decay
through intermediate states involving $K^*K$, where $K^*$ may be
$K^*(890)$, $K_1(1270)$ or $K_1(1400)$ [38].

\begin {figure}[t]
\begin {center}
\epsfig{file=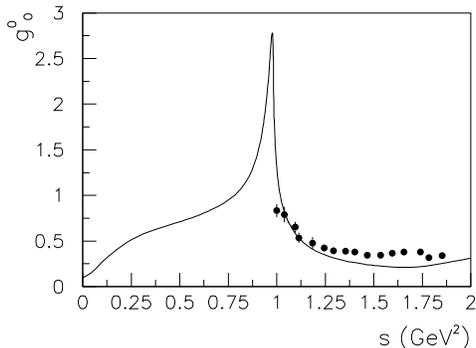,width=8.5cm}\
\vskip -8mm
\caption{The magnitude of the $\pi \pi \to KK$ amplitude from
B\" uttiker et al [39] fitted using the Roy equations and
experimental data of Cohen et al [40].}
\end {center}
\end {figure}

A further alternative is proposed here.
This is that the broad $\sigma$ couples to the $KK$ loop.
It will be shown in Section 5 that data on $\pi \pi \to KK$
require this broad component.
Independent support is available from the work of
B\" uttiker, Descotes-Genon and Moussallam [39].
They apply the Roy equations to fitting data on $\pi K \to \pi K$
and $\pi \pi \to KK$.
For the present discussion, this corresponds to
calculating the $\pi \pi \to KK$ amplitude below the $KK$ threshold
as a continuation of the amplitude above threshold
using \linebreak (i) the dispersion integral over the physical
$\pi \pi \to KK$ process, (ii) nearby pole terms due
to $t$ and $u$-channel exchanges of $\rho (770)$
and $K^*(890)$.
Their Figure 12, reproduced here as Fig. 5,
shows the magnitude $|g^0_0|$ of their $\pi \pi \to KK$ amplitude both
above and below the $KK$ threshold. The $f_0(980)$
signal is superimposed on a strong broad component.
Achasov and Kiselev [29] similarly include coupling of $\sigma$ to
$KK$, but they fit parameters of $f_0(980)$ freely.

\begin {figure} [htb]
\begin {center}
\epsfig{file=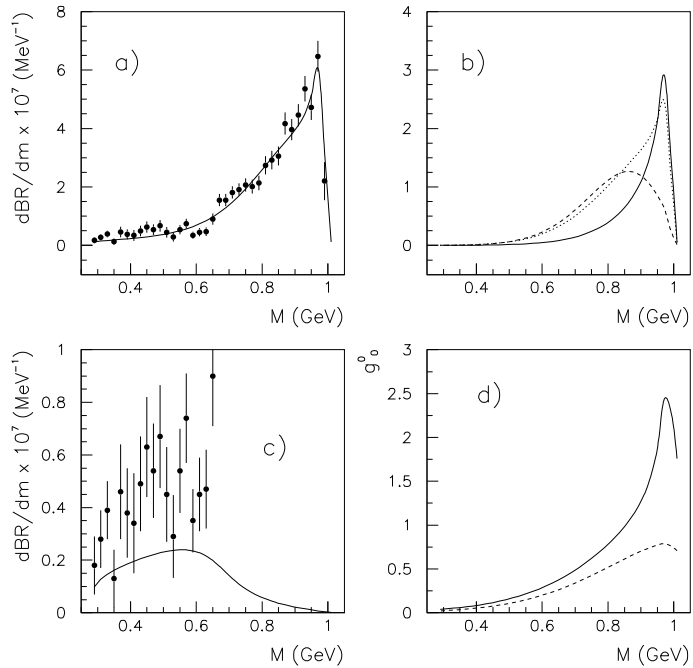,width=10cm}\
\caption{(a) Fit to KLOE $\gamma \pi ^0 \pi ^0$ data after
background subtraction; (b) contributions from $f_0(980)$ (full
curve), $\sigma$ (dashed) and interference (dotted); (c)
an enlarged view of the contribution from $\rho \pi$, $\rho \to
\gamma \pi ^0$; (d) $|g^0_0|$ from my fit (full curve) and the
contribution from $\sigma$ alone (dashed).}
\end {center}
\end {figure}

For the fit presented here,
the $f_0(980)$ contribution is fixed in magnitude to BES parameters.
The charge form factor of the photon to $KK$ is fitted freely.
The magnitude of $\sigma \to KK$ optimises with
a threshold ratio $r_2 = g^2_{KK}/g^2_{\pi \pi} = 0.6 \pm 0.1$.
Its phase with respect to $f_0(980)$ requires some comment.
The production process is electromagnetic, and will produce
a phase shift only of order $\alpha = 1/137$.
In $\pi \pi$ elastic scattering, both $\sigma$ and $f_0(980)$
have a phase shift of $\sim 90^\circ$ at 990 MeV.
Their relative phase is constrained in fitting KLOE data
within $\pm 10^\circ$, the combined error
from $\sigma$ phases and the mass of $f_0(980)$.
It fits naturally at the edge of this band;
if the constraint is removed, $\chi ^2$ changes by only 2
and the fit hardly changes.

The resulting fit is shown in Fig. 6(a).
Panel (b) displays the
$f_0(980)$ intensity as the full curve, the $\sigma$
as the dashed curve and the interference between them as the dotted
curve.
Panel (c) shows the fitted $\rho \pi$ signal (full
curve) compared with data at low masses. Its magnitude is fixed
according to the arithmetic of Section 2.4.
A marginal improvement may be obtained by fitting it
freely, but there is no significant effect on the fitted $\sigma $
amplitude.

Fig. 6(d) shows $|g^0_0|$ from my fit to $\pi \pi \to KK$ as the full
curve.
The peak due to $f_0(980)$ is slightly lower than that in Fig. 5
because the $f_0(980)$ of BES is slightly broader than that of
B\" uttiker et al.
Secondly, the fit requires a fairly strong form factor to
cut off the lower sides of both $f_0(980)$ and $\sigma$.
It is parametrised as
$\exp {-8.4k^2_\gamma }$, where $k$ is in GeV.
Comments on the origin of this factor will be given in Section 4.1.

\begin {figure}  [htb]

\begin {center}
\epsfig{file=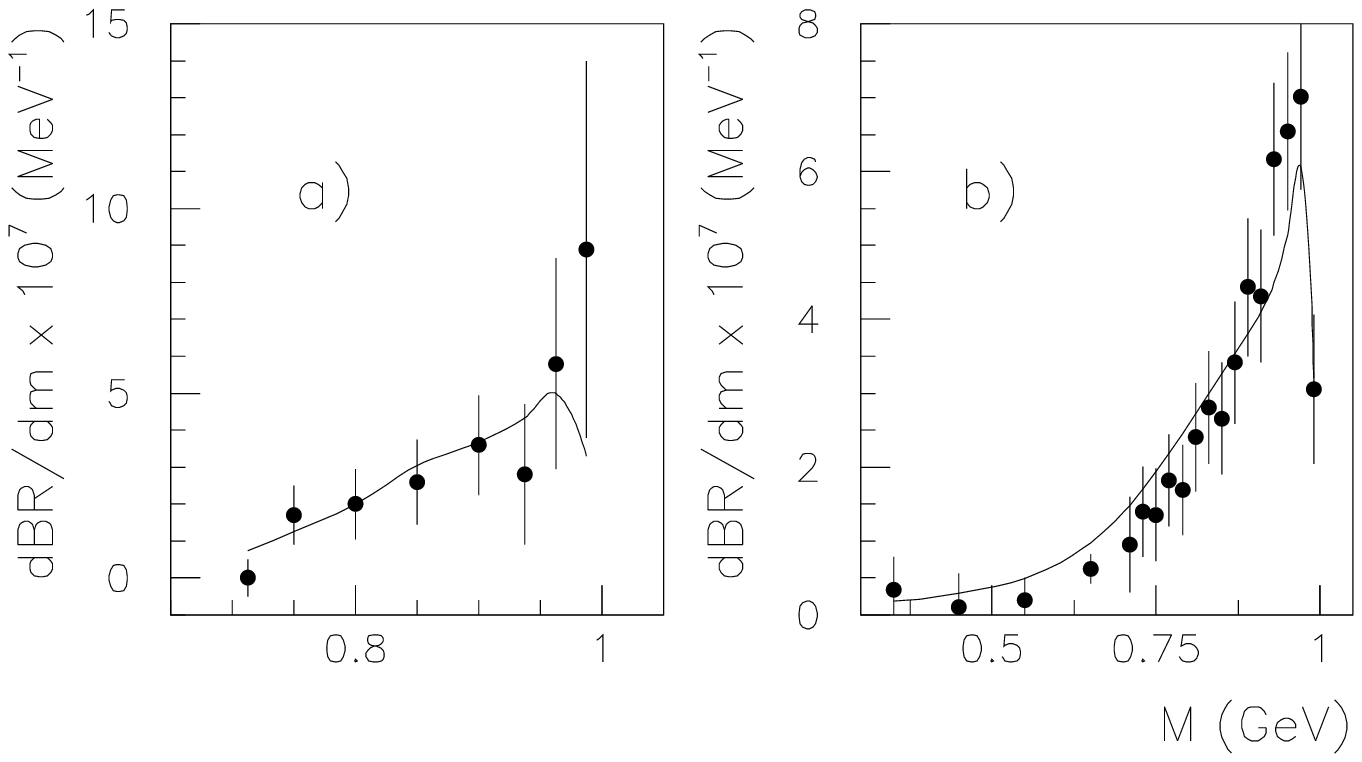,width=8cm}\
\vskip -10mm
\caption{Fits to Novosibirsk data on $\phi \to \gamma (\eta \pi ^0)$
[2] and $\phi (\pi ^0 \pi ^0)$ [3].}
\end {center}
\end {figure}

For completeness, Fig. 7 shows fits to Novosibirsk data on
$\phi \to \gamma (\eta \pi ^0) $ and $\gamma \pi ^0 \pi ^0$.
The point in $\gamma (\eta \pi ^0)$
at 0.98 GeV is clearly too high and is slightly inflating the PDG
average for the branching ratio $\phi \to \gamma (\eta \pi ^0)$.

\subsection {Sensitivity to fitting parameters}
The heights of the peaks at 970 MeV in both $\gamma (\eta \pi ^0)$
and $\gamma (\pi ^0 \pi ^0)$ have limited sensitivity
to resonance parameters.
In the denominators $D(s)$, $g^2_{\pi \pi }\rho _{\pi \pi}$ or
$g^2_{\eta \pi } \rho_{\eta \pi}$ appear in the imaginary part, while
$M$ and $g^2_{KK}\sqrt {4M^2_K/s - 1}$ appears in the real part.
There is considerable compensation between the numerator $N(s)$
and denominator $D(s)$ of the Breit-Wigner amplitudes.
For $\gamma (\eta \pi ^0)$, it is necessary to constrain the location
of the $a_0$ peak at half-height within, say, 2 or 2.5 standard
deviations of the values quoted in Section 3, adding statistical and
systematic errors linearly. With this constraint, $M$, $g^2_{\eta \pi}$
and $g^2_{KK}$ may be varied freely. Because of the feedback between
$N(s)$ and $D(s)$, it is only just possible to accomodate the 32\%
discrepancy by including also flexibility in the $\rho \pi$
contribution and uncertainty in the KLOE normalisation.

For $f_0(980)$, the BES $\pi \pi$ peak must again be constrained in
mass and the full width must be constrained to $34 \pm 4$ MeV.
There is more flexibility in $g^2_{KK}$, since the BES $f_0(980) \to
KK$ signal interferes with a background due to $\sigma \to KK$;
the error on $g^2_{KK}$, namely $\pm 0.25(stat) \pm 0.21(syst)$ quoted
by BES, represesents one standard deviation.
It turns out that with this flexibility, the absolute normalisation
of the $f_0(980)$ peak cannot change by more than $\pm 15\%$.
Some fits reported in the literature have changed $g^2_{\pi \pi}$ and
$g^2_{KK}/g^2_{\pi \pi}$ by factors of 2; that is unrealistic.

The peak at 970 MeV in KLOE data determines the magnitude of the
background amplitude there. The fit is not unduly sensitive
to the parametrisation of $\pi \pi$ phase shifts. What matters for
$\gamma (\pi ^0 \pi ^0)$ is only the way the $\sigma$
and $f_0(980)$ amplitudes go out of phase off resonance.
Any form which reproduces $\pi \pi$ phase shifts within errors of,
say, $\pm 4^\circ$ is adequate.
The full apparatus of eqns. (18)--(30) has been used to scrutinise
the combined fit with $\pi \pi \to \pi \pi$, $KK$ and $\eta \eta$.
An equivalent and more convenient form which fits all the data equally
well may be obtained by (i) dropping the dispersive contribution $m(s)$
completely, (ii) also the tiny contribution from the $4\pi$ channel
below the $KK$ threshold, (iii) replacing the effect of $m(s)$ for
$KK$ and $\eta \eta$ below their thresholds  by a form factor $\exp
(-5.2|k^2_K|)$, with $k_K$ in GeV/c. Then $M = 0.7366$ GeV,
$B_1 = 0.6470$ GeV$^2$, $B_2 = 4.271$ GeV$^2$, $B_4=0.00221$ GeV$^2$.
Values of other parameters will be discussed below but will
be collected here for convenience of reference:
$r_3=0.20$, $r_2=0.6$.

A crucial parameter in fitting $\sigma \to KK$ in KLOE data is the
form factor of Eqn. (24), $FF = \exp (-8.4k^2_\gamma)$, where $k_\gamma
$ is in GeV.
Without this factor, the background amplitude is too large
at low mass; one can see this in the difference between Figs. 5 and
6(d).
This form factor has at least three possible origins.

Firstly, it may be due to the size of the $KK$ cloud.
This would require an RMS radius of 1.4 fm.
It is well known that a source of finite size produces a
form factor $\sin (kr)/kr = 1 - k^2r^2/6 + \ldots$.
The second sheet pole of $f_0(980)$ lies at $(998 \pm 4) - i(17 \pm 4)$
MeV. The binding energy is so close to the $KK$ threshold that the
$f_0(980)$ inevitably has a $KK$ cloud of roughly this size, resembling
the long-range tail of the deuteron. The pole
for $a_0(980)$ is much further away, at $1032-i85$ MeV, so the RMS
radius of $a_0(980)$ has conventional dimensions and a weaker form
factor.

A second straightforward possibility is that $f_0(980)$ and
the broad $\sigma$ component mix over the mass range where
they are both large. This mixing will be confined naturally to
their region of strong overlap.

A third possibility is that $|g^0 _0|$ fitted by B\" uttiker et al.
has some flexibility in the range $s = 0.2-0.8$ fm.
It is well constrained near 1 GeV, and also for $s = -0.5$ to 0
GeV$^2$ (by data on $\pi K \to \pi K$).
In between, Adler zeros in $\pi \pi$ at
$s \sim 0.5 m^2_\pi$ and in coupling to $KK$ at $s \sim 0.5 m^2_K$
may pull $|g^0 _0|$ down. Data on $\phi \to \gamma (\pi ^0 \pi ^0)$
appear to be the only data directly sensitive to this possibility.

\section {Fits to $\pi \pi \to KK$ and $\eta \eta$}
\subsection {Introduction}
A problem with data on $\pi \pi \to KK$ is that cross sections
from different experiments differ in normalisation by up to a factor 2.
This problem is minimised here by using the BES parameters
for $f_0(980)$ to assist the absolute normalisation.
All $\pi\pi \to KK$ data are rescaled to the BES normalisation.
However, one can still not rely on $\pi \pi \to KK$ data to separate
branching fractions of $f_0(1370)$, $f_0(1500)$ and
$f_0(1790)$  to $KK$ and $\eta \eta$.
Vastly better determinations are available from the analysis of Crystal
Barrel data on $\bar pp$ annihilation at rest to $3\pi ^0$,
$\eta \pi ^0 \pi ^0$, $\eta \eta \pi ^0$ and several $KK\pi$ charge
combinations.
All  of these data have been fitted self-consistently by Sarantsev
and collaborators; decay widths to $\pi \pi$ and $\eta \eta$ are taken
from this analysis [41].
Their errors are used to set limits within which parameters
may vary; Table 1 shows these limits and values
at which they optimise.
Note however, that the present data really only determine the products
$\Gamma _{\pi \pi }\times \Gamma _{KK}$ and
$\Gamma _{\pi \pi }\times \Gamma _{\eta \eta}$.
Values of $\Gamma _{4\pi}$
are determined from $\Gamma _{4\pi} = \Gamma _{total} - \Gamma _{\pi\pi}
- \Gamma _{KK} - \Gamma _{\eta \eta }$ at each resonance mass.

\begin{table}[htb]
\begin {center}
\begin{tabular}{cccc}
\hline
State & $f_0(1370)$ & $f_0(1500)$ & $f_0(1790)$ \\\hline
$M$ &    1290--1335 (1290) & 1495--1510 (1501)
& 1770--1800 (1800) \\
$\Gamma_{total}$ & 200--280   (280)  & 110--135  (135) & 180-280 (280)
\\
$\Gamma_{\pi\pi}$ & 34-58   (48)  & 35--39  (35) & 30--100 (72)
\\
$\Gamma_{KK}/\Gamma _{\pi\pi}$\ & 0.04--0.235 (0.235)  & 0.22--0.272
(0.272) &  0.16--1.2 (0.47) \\
$\Gamma_{\eta\eta}/\Gamma _{\pi\pi}$\ & 0.08--0.18 (0.18)  &
0.04--0.18 (0.11) &  0.1--1.0 (0.20) \\\hline
\end{tabular}
\caption{Limits used in the fit (in MeV); optimised values are in
parentheses.} \end {center}
\end{table}

Some comments are needed on $f_0(1370)$ and $f_0(1790)$.
There has been some unnecessary controversy recently
concerning the existence of $f_0(1370)$, so its properties
will be outlined briefly; it will be the subject of a
separate publication giving full details from latest analyses
of Crystal Barrel data.
It appears most clearly in $\bar pp \to 3\pi ^0$,
where it is statistically at least a 50 standard deviation signal.
Statistics are 600K events, with almost no experimental
background. In early analyses of these data,
there was some sensitivity to the parametrisation of the $\pi \pi$
S-wave amplitude, with which it interferes.
The behaviour of the broad S-wave amplitude is now known much
better.

The $f_0(1370)$ decays dominantly to $4\pi$ with
$\Gamma (\pi \pi)/\Gamma (4\pi ) \le 0.2$.
The $4\pi$ inelasticity increases rapidly with mass, so it is
essential to treat the $s$-dependence of this width fully.
The result is that the $\pi \pi$ signal peaks at $1320 \pm
30$ MeV, as in the analysis of Anisovich et al. [41].
The $4\pi$ channel peaks approximately 75 MeV higher, because
of the rapidly expanding phase space.
The apparent width in $\pi \pi$ and $4\pi$ channels is likewise
different.
PDG listings assign unreasonably large errors because
these facts have not been taken into account in most
analyses.
The $s$-dependence of the $4\pi$ width is taken into account here,
including the associated dispersive correction to $m(s)$ in the
Breit-Wigner amplitude.

In elastic scattering, the intensity of $f_0(1370)$ is $<4\%$
of the unitarity limit. It is swamped by the $f_2(1270)$, which
has a spin multiplicity 5 times larger.
Achasov and Shestakov point out a rather large error in
$\Gamma _{4\pi}$ for $f_2(1270)$ [42].
This adds to the difficulty of normalising Cern-Munich moments in
the region of $f_0(1370)$.

The $f_0(1790)$ is required as well as $f_0(1710)$ by recent
BES data.
The $f_0(1710)$ appears clearly in $J/\Psi \to \omega K^+K^-$ [43].
It is conspicuously absent in $J/\Psi \to \omega \pi ^+ \pi ^-$
[36].
Those two sets of data require $\pi \pi/KK < 0.11$ for
$f_0(1710)$ with $95\%$ confidence.
In $J/\Psi \to \phi \pi ^+\pi ^-$, there is a distinct $\pi
\pi$ signal at 1790 MeV, but no significant signal in
$\phi K^+K^-$ [10].
The branching fraction $\pi \pi/KK$ is a factor 25 larger
than allowed for $f_0(1710)$, hence requiring a
separate resonance $f_0(1790)$.
There was previous evidence for it in
$J/\Psi \to \gamma 4\pi$ [44,45] and $\bar pp \to \eta \eta \pi ^0$
[46].
Here, both $f_0(1710)$ (with its
upper limit for $\pi \pi$ coupling) and $f_0(1790)$ (with its upper
limit for $KK$ coupling) are included.
However, the present data do not
distinguish cleanly between them.

\subsection {$\pi \pi \to \eta \eta$}
Fig. 8 shows fits to $\pi \pi \to KK$ data and $\eta \eta$.
The fit to $\eta \eta$ in panel (d) will be discussed first,
since it is simple and concerns only one set of data, from
GAMS  [47].
The normalisation is such that the unitarity limit
is at an intensity of 0.25; this arises from the relation
$T_{12} = S_{12}/2i$ between  T and S-matrices.
The full curve shows the overall fit, including contributions
from $\sigma $ (dotted curve), $f_0(1370)$, $f_0(1500)$ and
$f_0(1790)$/$f_0(1710)$.
The main features are a gradually falling $\sigma \to \eta \eta$ signal
and destructive interference with $f_0(1500)$, which is mainly
responsible for the dip at 1.45 GeV.
The peak at higher masses comes from $f_0(1500)$ interfering with
$f_0(1710)$ and $f_0(1790)$.
The Argand diagram is shown in Fig. 9(b).

\begin {figure} [htb]
\begin {center}
\epsfig{file=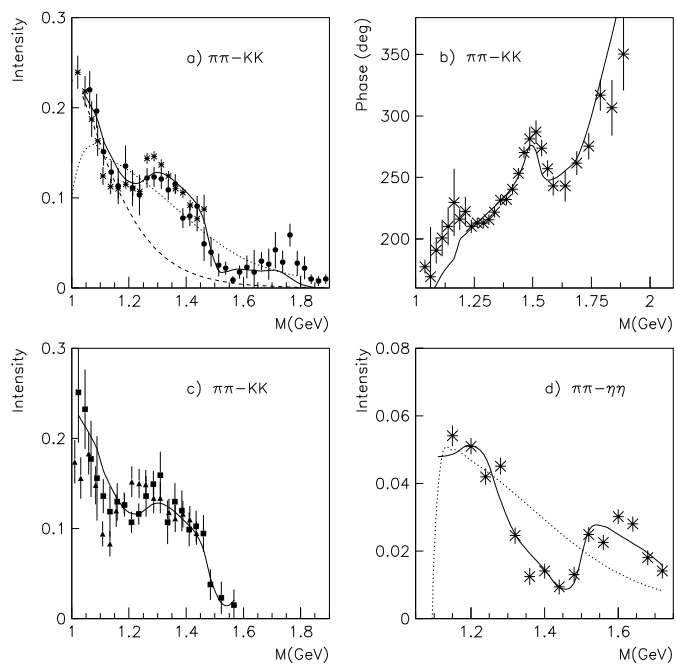,width=10cm}\

\caption{(a) Fit to $\pi \pi \to KK$ (full curve); circles show
data of Lindenbaum and Longacre [48] and crosses those of Martin and
Ozmutlu [52]; the dashed curve shows the intensity of $f_0(980)$
and the dotted curve that from $\sigma$; (b) fit to phases of
Etkin et al [54]; (c) fit to the data of Cohen et al. [40] (squares)
and Polychronakos et al. [53] (triangles);
(d) fit to data of Binon et al [47] for $\pi \pi \to \eta \eta$
(full curve); the dotted curve shows the $\sigma$ contribution.
}
\end {center}
\end {figure}

The fitted ratio $r_3 = g^2_3/g^2_1$ at the $\eta \eta$ threshold is
$0.20 \pm 0.04$. The error arises from normalisation
uncertainty and a possible weak coupling of $f_0(980) $ to $\eta \eta$;
an upper limit on this coupling
is obtained from BES data on $J/\psi \to \phi \pi ^+ \pi ^-$ and
$\phi K^+K^-$, from the fact that no drop is observed at the $\eta
\eta$ threshold in the $\pi \pi$ mass spectrum.
This upper limit is $r_3 \simeq 0.33$.

\subsection {$\pi \pi \to KK$}
It is first necessary to review the many available
sets of data.
It is important to realise that an analysis of moments was
required in order to separate spins $ J=0$, 1 and 2 and small amounts
of higher $J$.
The papers make
clear that there is some cross-talk with the high mass tail
of $\rho (770)$ and with $f_2(1270)$, whose branching ratio to $KK$ is
not  particularly well known.

\begin {figure}
\begin {center}
\epsfig{file=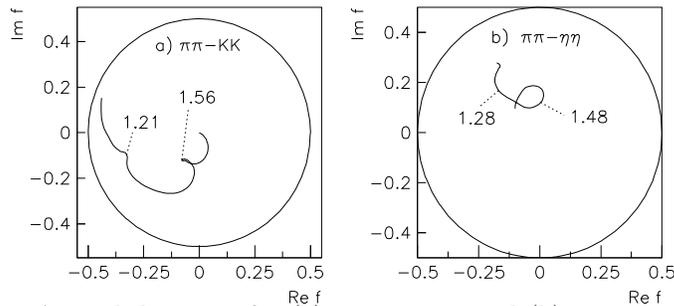,width=10cm}\
\vskip -10mm
\caption{Argand diagrams for (a) $\pi \pi \to KK$ and (b)
$\pi \pi \to \eta \eta$; masses are marked in GeV.
}
\end {center}
\end {figure}

Costa et al. [50] produced data on $\pi ^-p \to K^+K^-n$,
but found 8 alternative solutions, because of ambiguities
over the exchange process and the possibility of both $I=1$
and 0 contributions to the final state.
Cason et al. [51] produced data on $\pi ^- p \to K^0_SK^0_S$
and observed a shoulder at 1300 MeV.
They obtained two solutions, one with a narrow S-wave resonance
(width $\sim 95$ MeV) at this mass and the second with a more slowly
varying amplitude similar to today's $f_0(1370)$.
The situation clarified further with data from
Pawlicki et al. [49] on both $\pi ^- p \to K^+K^-n$ and $\pi ^+n \to
K^-K^+p$.
These data favoured the second solution of Cason et al. and
assigned $I = 0$.
A meticulous analysis of the $t$-dependence of data was made
by Martin and Ozmutlu [52].
This analysis showed that $\pi$ exchange dominates
at low $t$, as expected.
The paper of Polychronakos et al. [53] on $\pi ^- p \to
nK^0_SK^0_S$ is a full length paper on the data of Cason et al.
The publication of Cohen et al. [40]
gives a revised partial wave analysis of the data of Pawlicki et al.
and again favours a structure close to today's $f_0(1370)$.
Etkin et al. [54] reported data on $\pi ^- p \to K^0_SK^0_S n$.
Further statistics were added in the paper of
Lindenbaum and Longacre  [48].

It is noteworthy that all the later sets of data [48], [40] and
[53] observed a definite small bump at 1300 MeV, which was
christened the $\epsilon$ at the time and is now $f_0(1370)$.
All also observed a threshold peak  due to $f_0(980)$.
However, there are discrepancies
concerning the height of the $f_0(980)$ peak with respect to
the 1300 MeV mass range.
The analyses of Martin and Ozmutlu and Longacre et al. agree
well.
The revised analysis by Cohen et al. of the Pawlicki et al.
data gives a distinctly lower $f_0(980)$ than that of
Longacre et al.
And the data of Polychronakos et al. give an even lower
$f_0(980)$ peak.

To clarify the situation including today's information
about $f_0(1370)$, the four sets of data from Lindenbaum and
Longacre,
Cohen et al., Martin and Ozmutlu and Polychronakos et al.
were included into the inital stages of the analysis reported here.
This was despite the fact that those of Martin and Ozmutlu and
Cohen et al. are based on
the same actual data before the moments analysis.
The objective is to see whether the analysis favours one
or the other.
This is not the case and the fit goes midway between the two, so
they are both retained in the final fit, but with a weight half
that of Longacre et al.

The final fit is shown in Figs. 8(a) and (b).
The $f_0(980)$ produces the threshold peak, but
its tail at high mass cannot account for the remaining features.
Some $\sigma \to KK$ is definitely required and the dotted curve
shows the optimum fit.
It requires a form factor $FF=\exp (-5.2k^2)$ with $k$ is GeV/c.
However, some of this form factor may reflect the effect of
competition between $KK$ and $4\pi$ channels.
The $\eta \eta$ threshold contributes a drop in the
intensity of 0.02 between 1.09 and 1.18 GeV.
The $f_0(1370)$ helps fit the bump at 1300 MeV.
Destructive interference with $f_0(1500)$
is required to fit the dip at 1550 MeV.
The fit to the 1550--1900 MeV intensity is not perfect because of
difficulties in reproducing the large phase variation reported by
Etkin et al. [54].
The fit to their phases is shown in panel (b).
However, one should be aware that other phase determinations
show systematic differences with their data: the whole curve
can move up or down by up to $30^\circ$, although the trend
of the phase with mass remains similar to Fig. 8(b).
These differerences probably arise from the way spin 2 is fitted
in different analyses.
Etkin et al. included $f_2'(1525)$, but at that time the
existence of $f_2(1565)$ was not known, and could lead to
small systematic shifts.
Another possible source of systematic error is
that the present fit ends at 1900 MeV; there may be some contribution
from the known $f_0(2020)$ which is presently not fitted.

\begin {figure} [htb]
\begin {center}
\epsfig{file=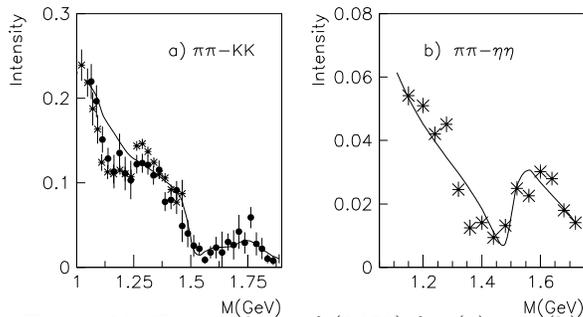,width=9cm}\
\vskip -10mm
\caption{Fits without $f_0(1370)$ for (a) $\pi \pi$, (b) $\eta \eta$.
}
\end {center}
\end {figure}

The optimum fit gives $r_2 = 0.6 ^{+0.1}_{-0.2}$, eqn. (22),
in close agreement with the KLOE data.
This value can depend systematically on the $s$-dependence fitted
to $f_0(980)$ and on $4\pi$ inelasticity.
The $f_0(980)$ is fitted with a conservative form factor $\exp
(-2.7k^2)$, where $k$ is kaon centre of mass momentum; this corresponds
to a radius of 0.8 fm for the $\pi \pi \to KK$ interaction,
i.e. a conventional radius.
Fig. 8(c) compares the fit with intensities derived by
Cohen et al [40] (squares) and those of Polychronakos et al. [53]
(triangles). The latter are systematically low below 1.15 GeV,
so they are omitted from the final fit.

Fig. 10(a) shows the fit without $f_0(1370)$. It
fails to fit the dip below 1300 MeV.
Since that structure is observed in 3 experiments and 4 sets of
data, it is hard to deny the presence of $f_0(1370)$; its fitted
mass and width agree within errors with the best determinations
from Crystal Barrel [41] and BES [10].
There is presently no alternative
explanation of the 1300 MeV bump. The onset of $4\pi$ inelasticity is
far too slow to account for structure with the width observed around
1300 MeV, and is anyway taken into account fully in the present
analysis.
Fig. 10(b) also shows the change in fitting $\pi \pi \to \eta \eta$
when $f_0(1370)$ is removed;
however, the points in Fig. 8(d) fluctuate around the fit by more
than statistics, so it is difficult to estimate the
significance of $f_0(1370) \to \eta \eta$.

\begin {figure} [htb]
\begin {center}
\epsfig{file=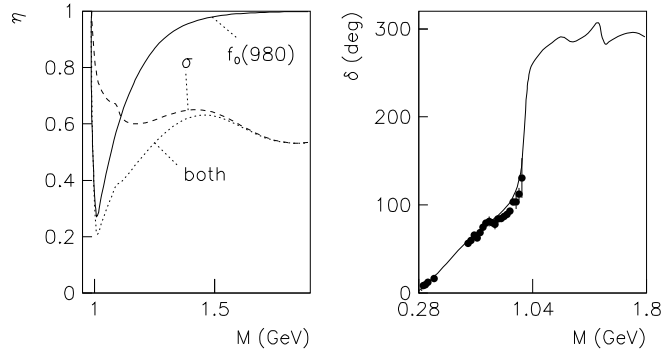,width=10cm}\
\vskip -8mm
\caption{(a) Contributions to elasticity $\eta$ for $\pi \pi$ elastic
scattering from $f_0(980)$ alone (full curve), $\sigma$ (dashed) and
their product (dotted). (b) Fitted $\pi \pi$ phase shifts;
points show data of Pislak et al. [55] below 400 MeV and Cern-Munich
data above 600 MeV [14]. }
\end {center}
\end {figure}

Fig. 11(a) illustrates the way elasticities of $f_0(980)$ and $\sigma$
combine. The dotted curve shows the result obtained by
multiplying $f_0(980)$ and $\sigma$ contributions, according to the
model assumed here. Fig. 11(b) shows the phase shift fitted to $\pi
\pi$ elastic scattering; errors above the $f_0(980)$ are typically $\pm
15^\circ$.

\section {Summary}
Data on $\phi \to \gamma (\eta \pi )$ agree within errors in both
absolute normalisation and mass spectrum with the parameters of
$a_0(980)$ from Ref. [8]; the KLOE data suggest a normalisation
which can be accomodated by shifting $a_0(980)$ parameter to
$M = 991.5 \pm 4$ MeV, $g^2_{\eta \pi} = 0.241 \pm 14$ GeV$^2$,
$g^2_{KK}/g^2_{\eta \pi} = 1.34 \pm 0.13$.
The errors quoted here are reduced compared with Ref. [8] after
the combined fit with KLOE data.
The agreement with KLOE data suggests that the $KK$ loop model
is reliable.
If BES parameters for $f_0(980)$ are used, data on $\phi
\to \gamma (\pi ^0 \pi ^0)$ have a normalisation higher than the fit by
at least a factor 2. This discrepancy may be solved by allowing
constructive interference between $f_0(980)$ and $\sigma \to KK$. The
fit requires a ratio $r_2 = g^2(\sigma \to KK)/g^2(\sigma \to \pi \pi
)= 0.6 \pm 0.1$ at the $KK$ threshold. Data on $\pi \pi \to KK$ also
require coupling of $\sigma$ to $KK$ with the same value of $r_2$
within experimental errors. Data on $\pi \pi \to \eta \eta$ require a
ratio $r_3 = g^2(\sigma \to \eta \eta)/g^2(\sigma \to \eta \eta )= 0.20
\pm 0.04$ at the $\eta \eta $ threshold.

The interpretation of branching ratios in terms of models for
$f_0(980)$, $a_0(980)$, $\sigma$ and $\kappa$ is discussed in an
accompanying publication.

\vskip 2mm
\noindent
{\bf Acknowledgements}
\newline
I am grateful to Dr. S. Giovanella for explaining technical points
concerning KLOE data.
I am grateful to Prof. H. Leutwyler for extensive discussions of
how to fit the $\pi \pi$ S-wave amplitude with analytic functions.
I wish to thank Prof. A. Sarantsev for discussion about the
status and parameters of $f_0(1370)$.
I also wish to thank Dr. C. Hanhart for valuable discussion on formulae.

\end {document}